\documentclass[sigconf, 10pt, nonacm=true, balance=false]{acmart}

\usepackage{listings}
\usepackage{color}
\usepackage{float}
\usepackage{multirow}

\definecolor{codegreen}{rgb}{0,0.6,0}
\definecolor{codegray}{rgb}{0,0,0}
\definecolor{codepurple}{rgb}{0.58,0,0.82}
\definecolor{backcolour}{rgb}{0.95,0.95,0.92}
 
\lstdefinestyle{mystyle}{
    backgroundcolor=\color{backcolour},   
    commentstyle=\color{codegreen},
    keywordstyle=\color{magenta},
    numberstyle=\tiny\color{codegray},
    stringstyle=\color{codepurple},
    basicstyle=\fontsize{8}{10}\ttfamily,
    breakatwhitespace=false,         
    breaklines=true,                 
    captionpos=b,                    
    keepspaces=true,                 
    numbers=left,                    
    numbersep=5pt, 
    showspaces=false,                
    showstringspaces=false,
    showtabs=false,                  
    tabsize=2,
    frame=single
}
 
\def\BibTeX{{\rm B\kern-.05em{\sc i\kern-.025em b}\kern-.08emT\kern-.1667em\lower.7ex\hbox{E}\kern-.125emX}}

\begin{document}

%
\title{Lightweight assistive technology: A wearable, optical-fiber gesture recognition system}

%
\author{Sanjay Seshan}
\affiliation{%
  \institution{Fox Chapel Area High School, Grade 11}
  \city{Fox Chapel}
  \state{PA}
}

%
\renewcommand{\shortauthors}{Seshan}

%

\begin{abstract}
The goal of this project is to create an inexpensive, lightweight, wearable assistive device that can measure hand or finger movements accurately enough to identify a range of hand gestures. One eventual application is to provide assistive technology and sign language detection for the hearing impaired. 

My system, called LiTe (Light-based Technology), uses optical fibers embedded into a wristband. The wrist is an optimal place for the band since the light propagation in the optical fibers is impacted even by the slight movements of the tendons in the wrist when gestures are performed. The prototype incorporates light dependent resistors to measure these light propagation changes. When creating LiTe, I considered a variety of fiber materials, light frequencies, and physical shapes to optimize the tendon movement detection so that it can be accurately correlated with different gestures. 

I implemented and evaluated two approaches for gesture recognition. The first uses an algorithm that combines moving averages of sensor readings with gesture sensor reading signatures to determine the current gesture. The second uses a neural network trained on a labelled set of gesture readings to recognize gestures. Using the signature-based approach, I was able to achieve a 99.8\% accuracy at recognizing distinct gestures. Using the neural network the recognition accuracy was 98.8\%. This shows that high accuracy is feasible using both approaches. The results indicate that this novel method of using fiber optics-based sensors is a promising first step to creating a gesture recognition system.

\end{abstract}

%

%
\maketitle

\section{Introduction}
\label{sec:intro}

Hand gesture-based user interfaces have captured the attention of both researchers and users for many years. Such interfaces have often appeared in science fiction, such as the 2002 movie \textit{Minority Report} and the more recent \textit{Iron Man} movies, to depict advanced and easy to use computing systems. The value of a hand gesture interface is that it is often much easier to communicate particular concepts using gestures rather than using words or other means. Hand gestures play an especially important role in communication and the exchange of ideas for the speech or hearing impaired. In addition to providing a user interface, a gesture interaction system could potentially translate sign language to text or spoken language for the user.

Over the past several decades, the computing community has worked towards making such gesture-based interfaces a reality. While much progress has been made, practical systems of this type remain only in the realm of science fiction.  The eventual goal of my research is to create a solution that would not just work in theory, but a practical solution that could be used by the hearing impaired to communicate with others who do not know sign language.
Existing designs are often expensive and rely on the user wearing cumbersome and awkward hardware. For example, many designs are based on wearing rings, gloves or using external cameras to identify gestures. 
Based on the limitations of current approaches, I set five main goals for my gesture-based interface system: 1) the system must be wearable; 2) it must be non-intrusive; 3) it must be lightweight; 4) it must be low-powered; and 5) it must be inexpensive.

My system, LiTe, involves a novel method of using optical fibers to detect movements in the tendons in the human arm. Each limb on the human body contains tendons that reach from each digit all the way to the forearm. When each digit in the hand is moved, there is a minuscule movement (<1mm) in the tendons in the arm. My design relies on optical fiber-based sensors embedded in a strap worn across the transverse carpal ligament where these tendons are located. As twenty percent of all American adults own a smart watch, these sensors could be incorporated into an instrumented watch strap providing a practical and less intrusive possibility for gesture recognition. 

In this paper, I address two broad categories of challenges in the design of LiTe:
\begin{itemize}
    \item {\bf Sensor Design.} The sensor must be sensitive enough to accurately detect even the minuscule movements of the tendons while also being in a form-factor that can be comfortably worn as a wristband. My work considers the material to create the fiber, the shape of the fibers and the wavelength of light used for sensing. 
    \item {\bf Gesture Detection.} LiTe must identify the gesture performed using the raw sensor values. I implement and evaluate two approaches for gesture recognition. The first uses an algorithm that combines moving averages of sensor readings with {\it gesture sensor reading signatures} to determine the current gesture. The second uses a neural network trained on a labeled set of gesture readings to recognize gestures. 
\end{itemize}

I built a prototype of the LiTe system using custom made Vytaflex fibers, Large 630nm (red) LEDs. Using the signature-based approach, I was able to achieve a 99.8\% accuracy at recognizing a set of 3 distinct gestures. Using the neural network the recognition accuracy was 98.8\%. This shows that high accuracy is feasible using both approaches. The neural network approach allows for more automation and the addition of gestures. The neural network accuracy can be further improved using alternative neural network architectures. The data shows that the LiTe design is capable of accurately identifying gestures using simple and unobtrusive hardware. It provides a promising first step to a practical gesture based system that can be used in a multitude of applications including gaming, robotics, biomedical engineering, especially offering improved communication for the speech and hearing impaired.

The remainder of the paper is organized as follows. In Section~\ref{sec:related}, I describe existing gesture recognition systems and background relevant to my design. Section~\ref{sec:materials} discusses how optical fibers can be used to detect movement. I describe my prototype hardware and hardware testing in Section~\ref{sec:prototype}. Section~\ref{sec:software} presents the software components used for the prototype. Section~\ref{sec:results} details the results and data analysis.

\begin{figure}[t]
  \centering
  \includegraphics[width=\linewidth]{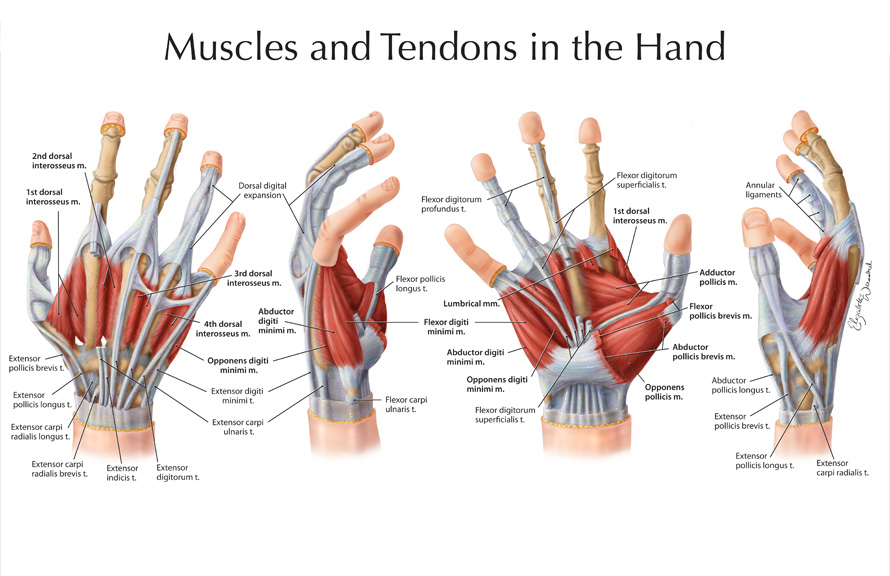}
  \caption{Tendons in the hand and wrist. Image Source:~\cite{medicalart}} 
  \label{fig:tendons}
\end{figure}

\section{Related Work and Background}
\label{sec:related}

\subsection{Existing Designs}
There are currently two types of designs used in gesture recognition systems: 1) solutions that are wearable, such as rings or gloves; and 2) solutions that require external cameras and radar systems.

One common solution is to use rings on fingers or smart gloves to detect movement. For instance, Tap~\cite{tap} is a wearable keyboard. It uses a system with five interconnected rings on a hand. Each ring contains an accelerometer that is used to identify distinct hand gestures. Tap costs \$150. Cyberglove~\cite{cyberglove} uses a glove with sensors designed to determine the angle of each joint. Cyberglove costs roughly \$15000 to \$30000.  These systems are intrusive since humans do not  to wear gloves nor sets of rings all day.

Other designs rely on the use of deployed infrastructure to detect gestures. For example, Microsoft's Xbox Kinect system~\cite{kinect} relies on the use of depth-sensing cameras to identify body pose and movement. Other designs such as the Leap Motion Controller~\cite{leap} use standard cameras to determine the gesture made. Google's Pixel 4 phone~\cite{pixel4} uses an embedded radar based system to detect simple hand gestures. Both camera and radar systems are limited by the ability to use gestures in a limited area -- e.g. in front of the Kinect camera or above the phone radar. My goal is to design a gesture recognition system that is less constrained in its usage. 

\subsection{The Biology}

The basis of LiTe's design lies in the biology of how finger movement uses tendons. Tendons are essentially fibrous cords. They are bands of connective tissue that attach the muscles to the bone enabling the muscles to move the bones. The main tendons of the hand (Figure~\ref{fig:tendons}) are as follows ~\cite{biology}: 
\begin{itemize}
    \item {\bf Superficialis.} These tendons run through the palm, attach at the bases of the middle phalanges, and are used to flex the wrist and finger joints.
     \item {\bf Profundus.}  These tendons pass through the palm, attach at the bases of the distal phalanges, and are used to flex the wrist and finger joints.
    \item {\bf Extensor.} These tendons are found in the fingers, which attach to the middle and distal phalanges, and extend the wrist and finger joints.
    \item {\bf Flexor.} These nine tendons pass from the forearm through the carpal tunnel of the wrist. At the palm, two go to each finger and one also goes to the thumb.
     \item {\bf Extensor pollicis brevis and Abductor pollicis longus.} These run from the muscles in the top of the forearm and allow the thumb to move.
\end{itemize}

 LiTe's primary design feature is to detect movements in these tendons.
\section{Optical Fibers for Gesture Recognition}
\label{sec:materials}

LiTe relies on the use of optical fibers to create a lightweight compact sensor for detecting tendon movement.

\begin{figure}[t]
  \centering
  \includegraphics[width=\linewidth]{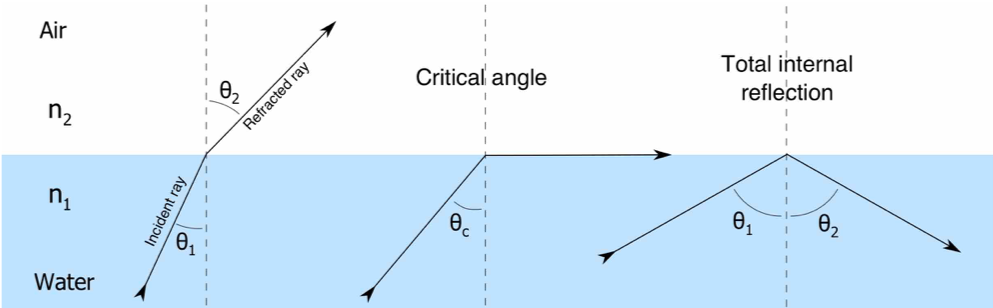}
  \caption{Total internal reflection. Image Source:~\cite{physics}}
  \label{fig:critcal}
\end{figure}

Before diving into the design of LiTe, we must understand how they can be used as sensor to detect deflection. Optical fibers rely on the concept of total internal reflection (all the light is reflected back into the fiber) to cause light to travel from one end of the fiber to the other (Figure~\ref{fig:critcal}). This is caused by the differences in the refractive indices between the fiber medium and the surrounding area. The refractive index of a material is determined by the following equation, where c is the speed of light:
\[ n=c_{medium}/c_{vacuum} \]
Inside the fiber itself the light reflects at the angle
\[ \theta_c=arcsin(\frac{n_{atmosphere}}{n_{medium}}) \]

For example, for most glass fiber optics, the following is true:
\[n_{medium} = 1.444 \]
\[n_{atmosphere}=1.000350 \]
\[\theta_c=43.8^{\circ}\]

\begin{figure}[t]
  \centering
  \includegraphics[width=0.5\linewidth)]{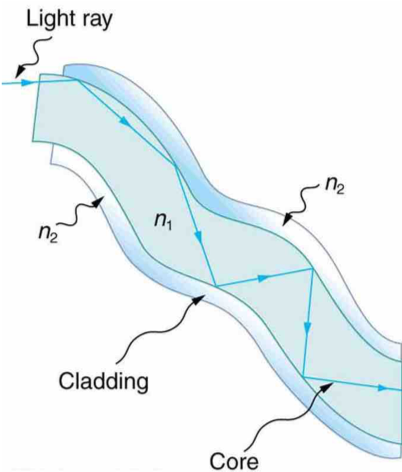}
  \caption{Light traversing a fiber. Image Source:~\cite{physics}}
  \label{fig:fiber}
\end{figure}

\begin{figure}[t]
  \centering
  \includegraphics[width=0.5\linewidth]{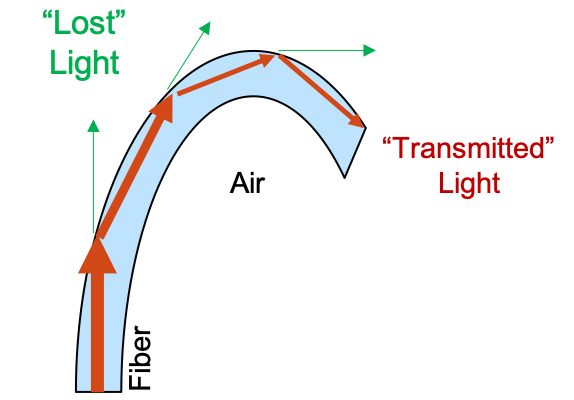}
  \caption{Light loss due to imperfections.}
  \label{fig:fiberloss}
\end{figure}

This allows light to travel from one end of the fiber to the other (Figure~\ref{fig:fiber}).  Ideally, no light is lost through the sides of the fiber, with all the light travelling between the ends. However, this system is not perfect and this project depends on the imperfections in order to detect the deflection in the arm. These imperfections may include a not high enough difference in refractive index between the medium and the atmosphere and physical inconsistencies in the material, such as air bubbles or semi-transparency. The idea is that these imperfections will cause some of the light to be lost as it travels down the fiber (Figure~\ref{fig:fiberloss}). Most importantly, the amount of light lost is significantly impacted by any deflection or bending of the fiber. My sensor design measures the amount of light that successfully traverses the fiber to measure tendon deflection.

\begin{figure}[t]
  \centering
  \includegraphics[width=\linewidth]{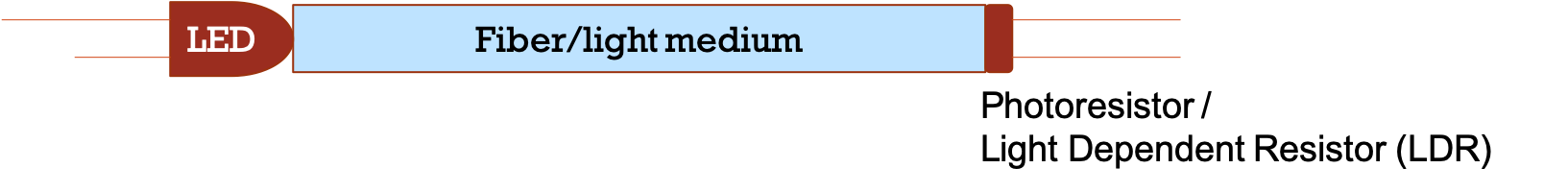}
  \caption{Basic Setup for Deflection Sensor}
  \label{fig:basic_setup}
\end{figure}

The basic setup is as follows. Most optical fibers have two layers: the medium itself and the cladding (outer covering). In this project, the cladding is simply Earth's atmosphere to maximize the loss of light across the fiber. An LED is connected to one end of an optical fiber using some heat shrink. An Light Dependent Resistor (LDR), or photoresistor, is fastened to the other end. The LDR is used to measure the light intensity across the fiber, and measures the amount of light that is transmitted when bent. (Figure~\ref{fig:basic_setup}) This setup provides the possibilities for detecting deflections of tendons in the wrist.


While the concept of using optics as sensors has been explored before by researchers, my application and manufacturing process is unique as explained in the next section. I leverage and improve upon the prior research in the area of optical fibers as sensors. For example, the project \textit{Highly stretchable optical sensors for pressure, strain, and curvature measurement }\cite{strechable} creates a sensor system using optical fibers to detect curvature. This, and similar projects, \cite{optoelectronic} focus on the application of optics as sensors in robotics. I apply this concept to humans and use my own methodology to create my Proof of Concept.

\section{Proof of Concept}
\label{sec:prototype}

While the concept of using fiber light loss to detect bends in the fiber works in practice, designing a prototype that can measure the tiny bends caused by tendon movement requires addressing several significant design challenges. In developing a viable prototype, I created a testbed in which I could evaluate the best material (medium) to make the fiber out of, the best shape of the fiber to use, and the best wavelength of LED for the objectives.

\subsection{Testbed Hardware}

\begin{figure}[t]
  \centering
  \includegraphics[width=\linewidth]{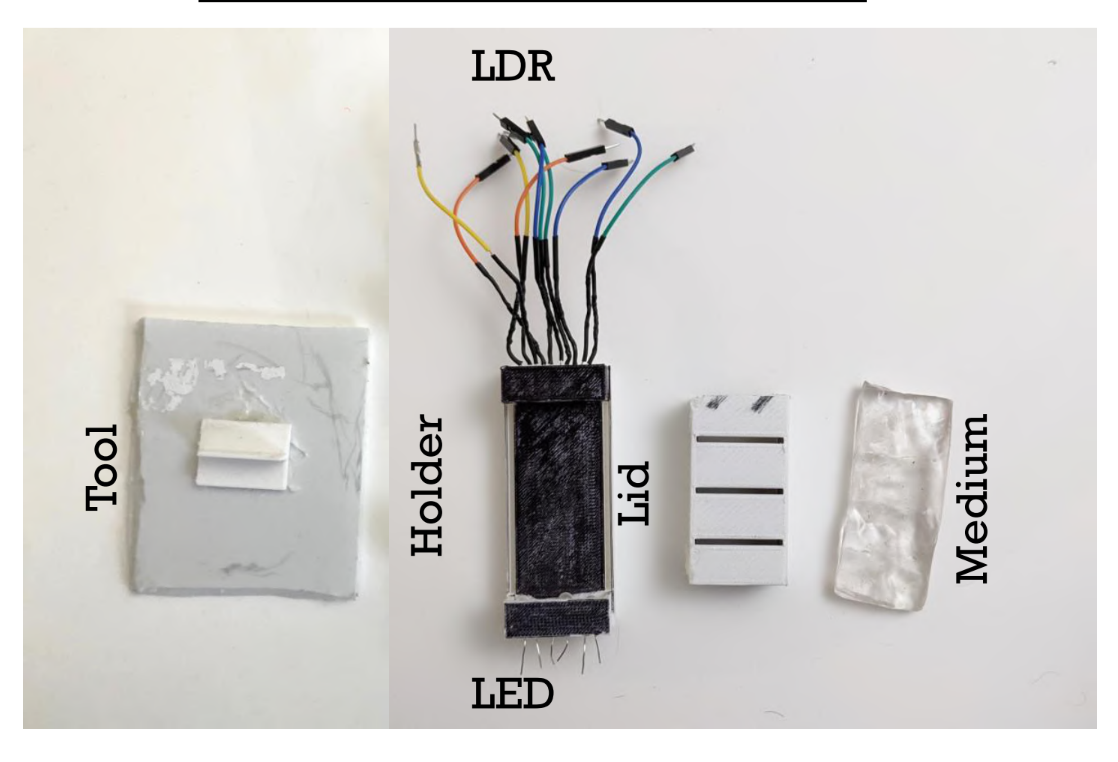}
  \caption{Testbed hardware}
  \label{fig:testbed}
\end{figure}

A testbed (Figure~\ref{fig:testbed}) was designed using Solidworks. Using this process, I created 1) a hard holder for the material to achieve both consistency and durability during the testing process, 2) a lid for to eliminate the external light and isolate the LED light, 3) a compartment in the holder for the LED and LDR on each end, and 4) a plastic tool for consistent depression of the medium.

\subsection{Optimal Wavelength}

\begin{figure}[t]
  \centering
  \includegraphics[width=\linewidth]{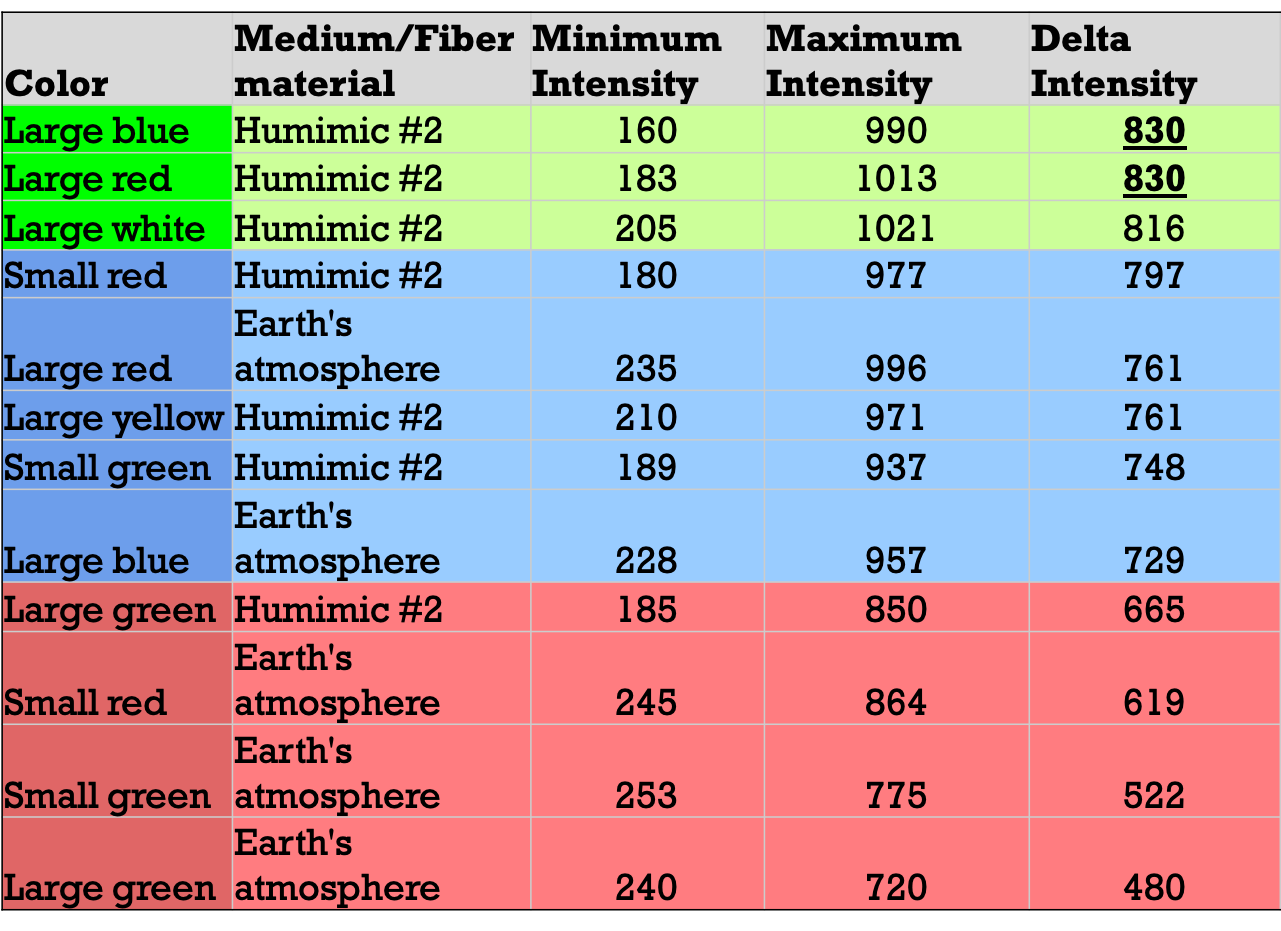}
  \caption{LED test results}
  \label{fig:LED}
\end{figure}

Using the above testbed hardware, the first step was to determine the optimal wavelength for the Light Dependent Resistor (LDR). The optimal wavelength depends on the type of LDR used. To determine this wavelength, an experiment with the testbed was conducted, comparing different LED wavelengths to the change in light intensity when depressed (Figure~\ref{fig:LED}). The higher the delta intensity is, the better the LDR reacted to that wavelength. For my Cadmium Sulphide (CdS) LDR, large red (630nm) and large blue (460nm) LEDs were determined to be the best as they produced the greatest delta intensity when depressed. Red was chosen due to availability. All measurements are in a scaled light intensity (0 to 1023), which represents the voltage drop (0 to 5V) across the light dependent resistor.

\subsection{Optimal Medium}

\begin{figure}[t]
  \centering
  \includegraphics[width=\linewidth]{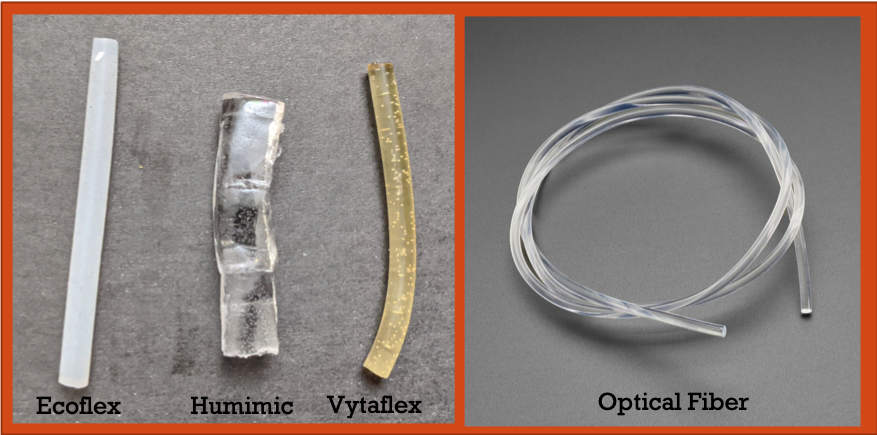}
  \caption{Different mediums tested}
  \label{fig:medium}
\end{figure}

The next step was to determine the best material of fiber to use. Logically, in order to be able to detect deflection over a small distance (i.e. the wrist), a flexible, soft, and mostly transparent material is required. From online research, the ideal types of materials that fit these properties are silicone, rubber, synthetic gelatin, and soft plastic. Therefore, I selected the following materials to test: 1) Smooth On Ecoflex 30, a silicone; 2) Humimic Gel 2, a synthetic transparent gelatin; 3) Smooth On Vytaflex 20, a rubber; and 4) standard optical fibre from Adafruit (shown left to right in Figure~\ref{fig:medium}).

Smooth On Ecoflex is an opaque silicone product. When testing the medium, it was found to be too opaque and failed to let any of the light from the LED through the material.  Humimic gel is a transparent synthetic gelatin. During testing, this medium failed since it is too soft. The medium melts even from body heat and would not be appropriate for use in a wrist band. Standard optical fiber (from Adafruit) did not work. The light intensity did not change with depression and was always at the
maximum (1023). Smooth On Vytaflex is a semi-transparent urethane rubber. This medium worked the best. It was able to transmit light and also maintained its structure. It also exhibited significant changes in intensity when depressed. 

\subsection{Optimal Fiber Shape}

\begin{figure}[t]
  \centering
  \includegraphics[width=\linewidth]{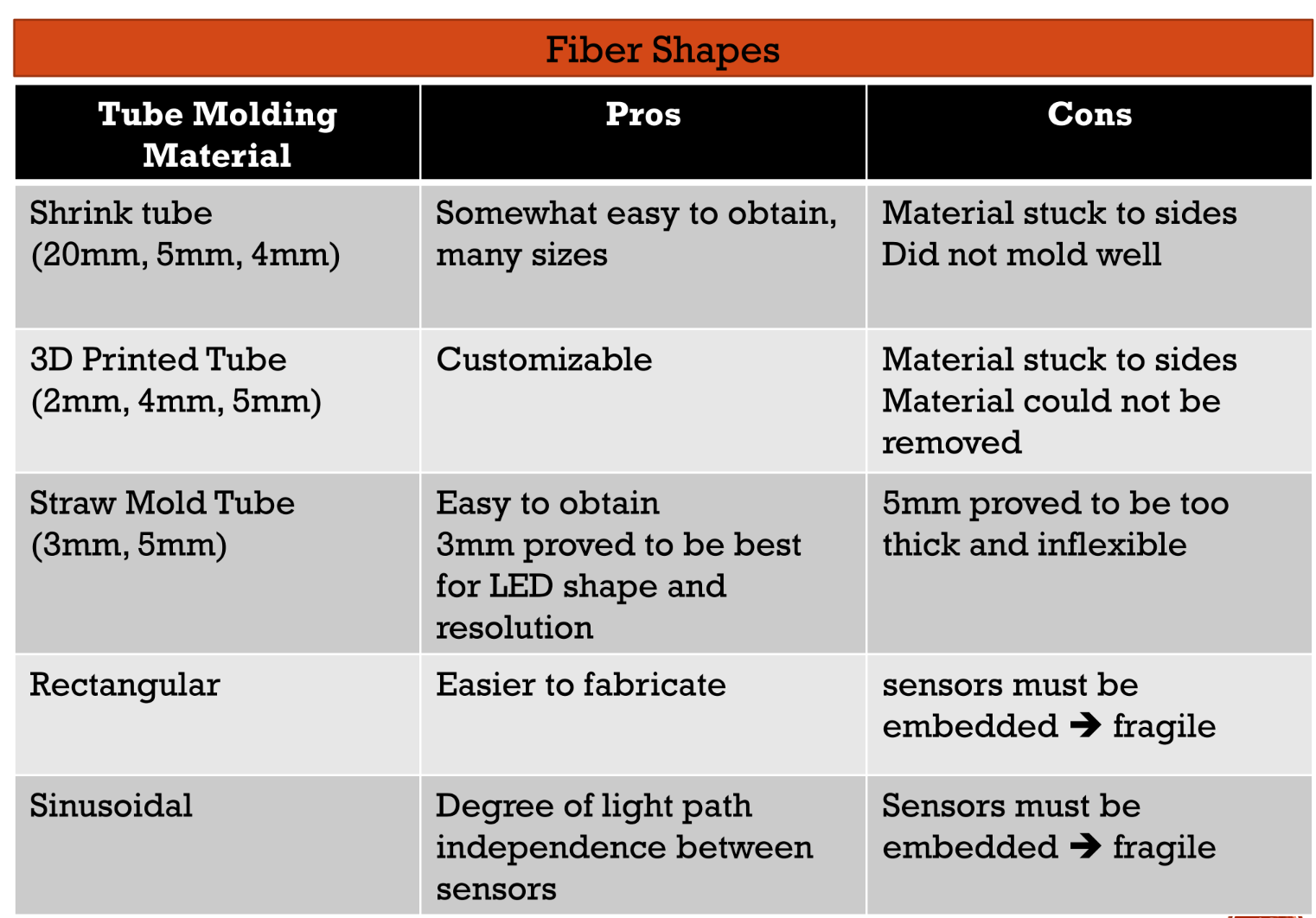}
  \caption{Molds and Shapes.}
  \label{fig:shape}
\end{figure}

The third step in the process was to determine the optimal shape of the fibers. To do this, I molded the medium using shrink tubes of various diameters, created a 3D printed tube, used straws and also created rectangular and sinuoidal molds (Figure~\ref{fig:shape}). The shrink tube molds were somewhat easy to obtain, but the Vytaflex had a tendency to stick to the sides. The 3D printed tubes were customizable as I was fabricating them myself, but the medium would stick to the sides and could not be removed. The rectangular and sinusoidal molds did create good shapes, but the sensors would have to be embedded into them, thus leaving them too fragile. In the end, the optimal shape ended up being a 3mm diameter straw mold. They were both easy to obtain and also the best for the LED shape and resolution used.

\subsection{Final Prototype Hardware}

\begin{figure}[t]
  \centering
  \includegraphics[width=\linewidth]{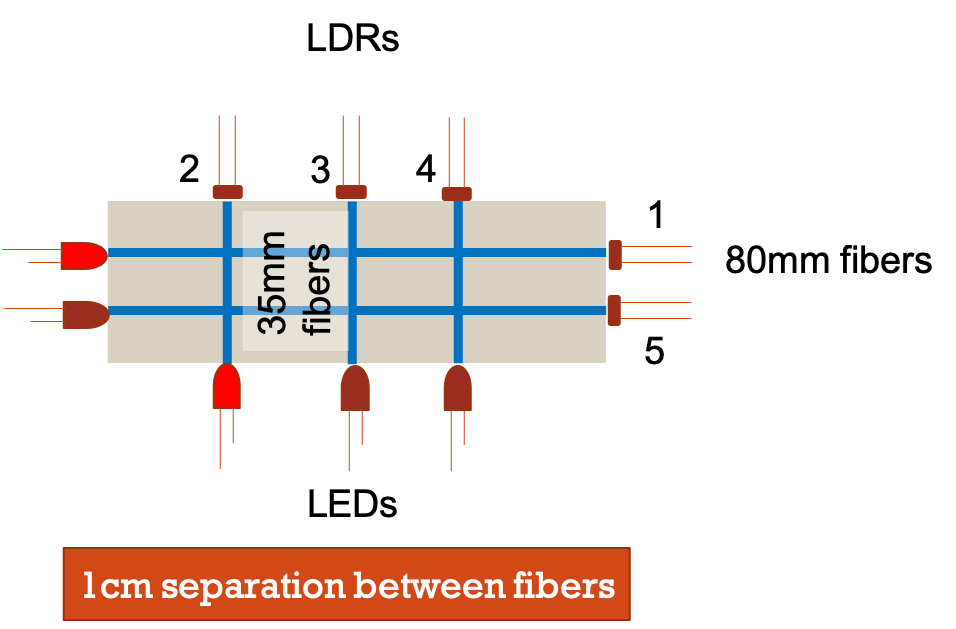}
  \caption{Prototype Layout}
  \label{fig:layout}
\end{figure}

The final prototype used in the project uses three 35mm long and two 80mm long Vytaflex fibers molded using 3mm straw. Large red (630nm) LEDs are used as the light source. The LEDs and LDRs are fastened using heat shrink to the fibers. (Figure~\ref{fig:layout})

\noindent Materials:
\begin{itemize}
\item Vytaflex fibers
\item Large red (630nm) LEDs
\item Arduino Uno
\item 5 LDRs and 5 10k Ohm resistors
\item LED circuit: 9V battery, 5V regulator, 50 Ohm resistor, 100 Ohm resistor, Ohms resistors, 5 LEDs

\end{itemize}

\begin{figure}[t]
  \centering
  \includegraphics[width=\linewidth]{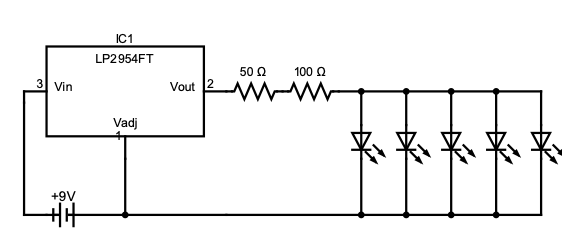}
  \caption{Circuit Diagram for LEDs}
  \label{fig:arduino}
\end{figure}

\begin{figure}[t]
  \centering
  \includegraphics[width=\linewidth]{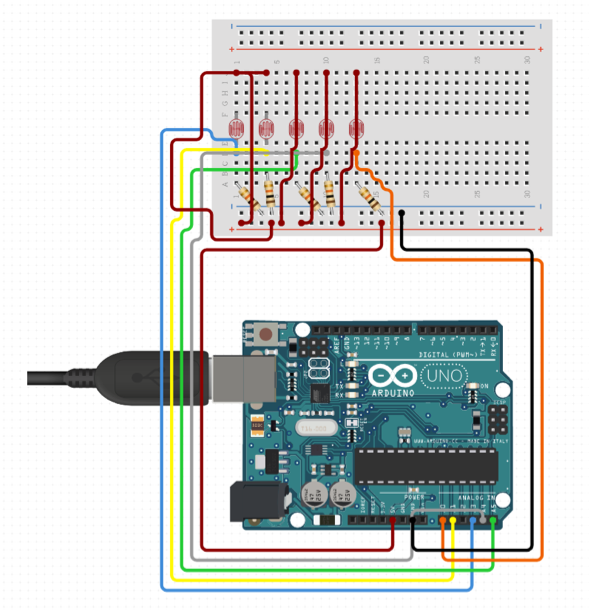}
  \caption{Arduino Circuit Diagram}
  \label{fig:circuit}
\end{figure}

\begin{figure}[t]
  \centering
  \includegraphics[width=\linewidth]{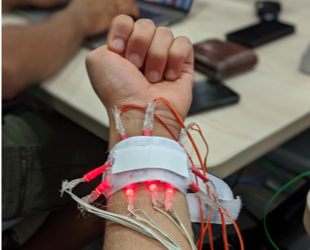}
  \caption{Final Prototype Hardware on Wrist}
  \label{fig:wrist}
\end{figure}

Figure~\ref{fig:arduino} and Figure~\ref{fig:circuit} depict how the prototype electronics was laid out. The grid-like system is used to be able to detect deflection both vertically and horizontally across the wrist. The entire setup is strapped to the wrist as seen in Figure~\ref{fig:wrist}

\section{Software Design}
\label{sec:software}

This section provides an overview of the LiTe system software; some of the error checking and details have been removed from the code segments below to improve readability. 

The first portion of the code, written in C++, runs on the Arduino Uno itself. The code uses the analog input ports on the Arduino to read the voltage drop across the light dependent resistor. I used the USB serial interface to send the values to the computer with a comma delimiter.


\begin{lstlisting}[language=C++]
// select the input pin for LDR(s)
int sensorPin = A0; 
int sensorPin2 = A1; 
int sensorPin3 = A2; 
int sensorPin4 = A3; 
int sensorPin5 = A4; 

// variable(s) to store the value coming from the sensor
int sensorValue = 0; 
int sensorValue2 = 0;
int sensorValue3 = 0; 
int sensorValue4 = 0; 
int sensorValue5 = 0;

void setup() {
    //sets serial port for communication
    Serial.begin(9600); 
    pinMode(LED_BUILTIN, OUTPUT);
}
void loop() {
    // read the value(s) from the sensor(s)
    sensorValue = analogRead(sensorPin); 
    sensorValue2 = analogRead(sensorPin2);
    sensorValue3 = analogRead(sensorPin3);
    sensorValue4 = analogRead(sensorPin4); 
    sensorValue5 = analogRead(sensorPin5); 

    //prints the values coming from the sensors on the screen
    Serial.print(sensorValue); 
    Serial.print(","); 
    Serial.print(sensorValue2); 
    Serial.print(","); 
    Serial.print(sensorValue3); 
    Serial.print(","); 
    Serial.print(sensorValue4); 
    Serial.print(","); 
    Serial.println(sensorValue5); 

}
\end{lstlisting}

This second portion of code is the Python Code that runs on a Debian GNU/Linux PC. This purpose of this code is to collect the data along with a timestamp and save it to a file for further processing. The code works by opening the serial port of the Arduino and creating a list of the gestures to test. A timer is used so that the user completes one gesture cycle in five seconds. The collected sensor readings are given preliminary gesture identifying labels based on this gesture cycle. The data output in the file consists of a timestamp, each individual LDR reading, and the gesture cycle label. 

\begin{lstlisting}[language=Python]
#!/usr/bin/python

import serial
import time
import sys

# Determine the file to save data to
port = sys.argv[1]
fileOut = sys.argv[2]

f = open(fileOut, "w")

ser = serial.Serial(port, 9600, timeout=1)
def getData(timeout, label):
  timeI = time.time()
  while time.time()-timeI < timeout:
    try:
      reading = ser.readline()
      pr = [0,0,0,0,0,0,0,""]
      pr[0] = time.time()-timeIA
      pr[1] = time.time()
      pr[2] = int(reading.split('\n')[0].split(",")[0])
      pr[3] = int(reading.split('\n')[0].split(",")[1])
      pr[4] = int(reading.split('\n')[0].split(",")[2])
      pr[5] = int(reading.split('\n')[0].split(",")[3])
      pr[6] = int(reading.split('\n')[0].split(",")[4])
      pr[7] = label
      for i in pr:
        print str(i)+",",
        f.write(str(i)+",")
      print ""
      f.write("\n")
    except:
      pass

gestures = ["extend","fist","one"]*2

print("delta Time, Unix Time, ldr1, ldr2, ldr3, ldr4, ldr5, label")
f.write("delta Time, Unix Time, ldr1, ldr2, ldr3, ldr4, ldr5, label\n")

timeIA = time.time()
for gesture in gestures:
  continueQ = raw_input("Do "+gesture+"? [Y/n]: ")
  if continueQ == "y" or continueQ == "" or continueQ == "Y":
    getData(5.0, gesture)
  elif continueQ == "n" or continueQ == "N":
    continueQ = "n"
    
f.close()
\end{lstlisting}

There is an additional section of the code for the machine learning portion of this project. This code is discussed in Section~\ref{sec:results}.

\section{Experimental Results}
\label{sec:results}

\begin{figure}[t]
  \centering
  \includegraphics[width=\linewidth]{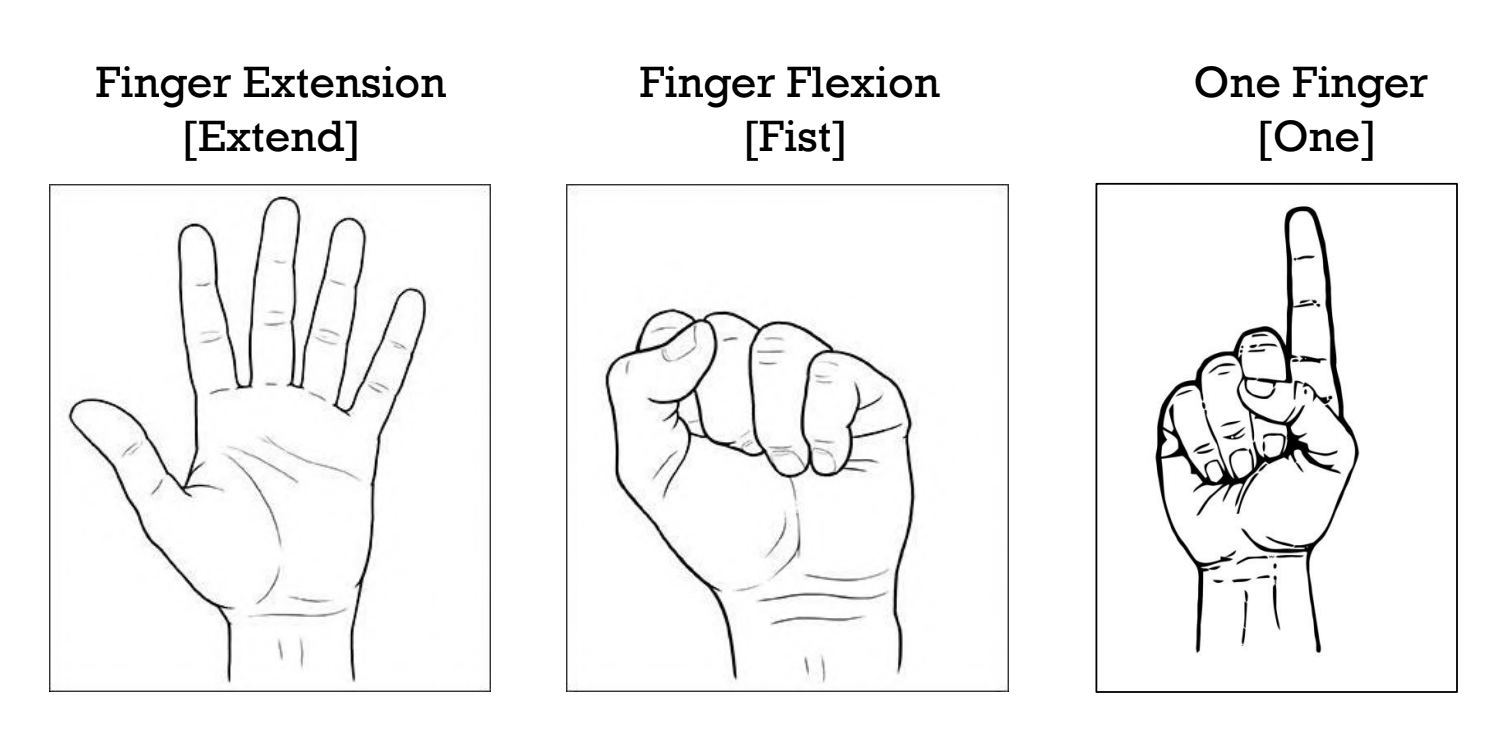}
  \caption{Gestures referenced in this paper }
  \label{fig:gestures}
\end{figure}

The goal of this project was to create a gesture recognition system. While a large collection of gestures can be tested using the prototype, I focus on three gestures, Finger Extension [EXTEND], Finger Flexion [FIST], and One Finger [ONE] (Figure~\ref{fig:gestures}), to evaluate in this paper to show the viability of the LiTe design. The key question I attempt to answer is whether the LiTe design can be used to accurately identify the gesture the user performs. 

To both train and evaluate my design, I produced a collection of datasets for both single gestures and mixes of different gestures. For each dataset, Light intensity readings from each LDR sensor was recorded once per 0.02s (250 readings/s). Light intensity measurement is reported in scaled units (0 to 1023) based on the voltage drop due to the LDR (0 to 5V). For single gesture datasets, the user repeatedly performed the gesture for approximately 5 seconds followed by a shorter (approximately 2 seconds) period during which the hand was relaxed. In the multiple gesture experiments, the user would cycle through the different gestures with a similar short relaxation period between each gesture. Each reading in each dataset was then manually labeled with the gesture being performed at the time. Five different data labels were used -- one for each gesture (EXTEND, FIST and ONE), one for the relaxed hand position (RELAX) and one for periods during which the hand was moving between two gestures (TRANSITION). The performance of LiTe is measured by how accurately it can predict these data labels using only the sensor readings. Note that I do not consider the accuracy of predicting the Transition label as part of the LiTe system performance since the Transition state may indeed reflect movement through other intermediate gestures. A practical deployment would use the stability of the gesture reading to eliminate this ambiguity. 

Below, I first describe the observations made on individual datasets that led to the design of a signature-based gesture classification algorithm (Section~\ref{sec:single}). I then describe the design and evaluation of the signature-based gesture classification (Section~\ref{sec:signature}). Finally, I describe and evaluate a neural network based classifier for gesture recognition (Section~\ref{sec:ML})


\subsection{Single Dataset Analysis}
\label{sec:single}

\begin{figure*}[t]
  \centering
  \includegraphics[width=\textwidth]{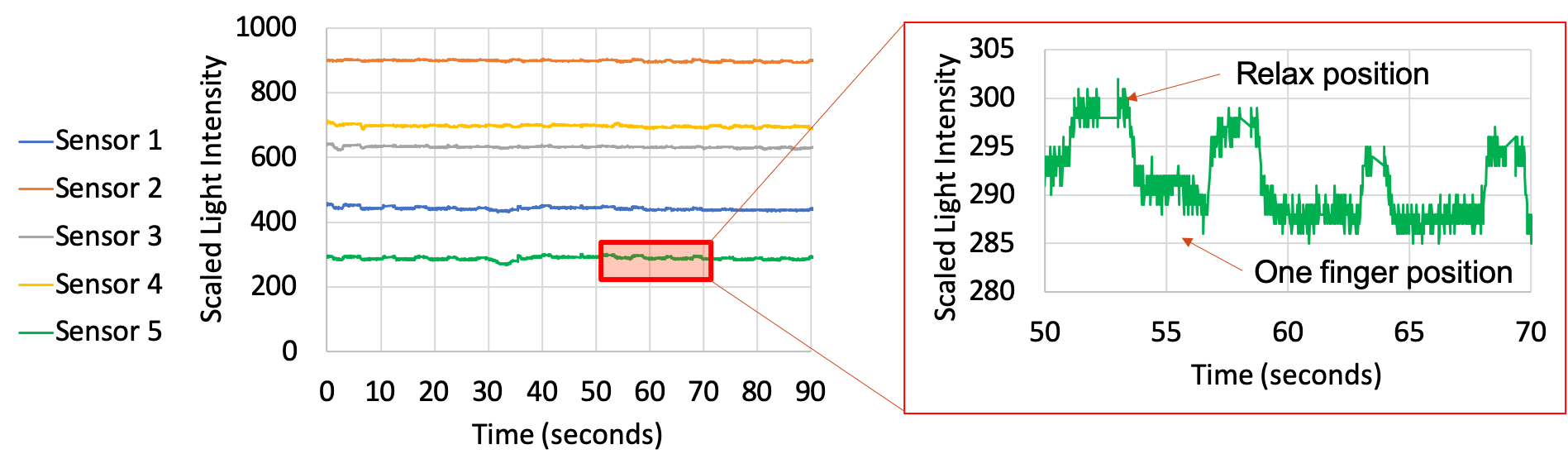}
  \caption{Sensor readings for the ONE gesture }
  \label{fig:one}
\end{figure*}

\begin{figure}[t]
  \centering
  \includegraphics[width=\linewidth]{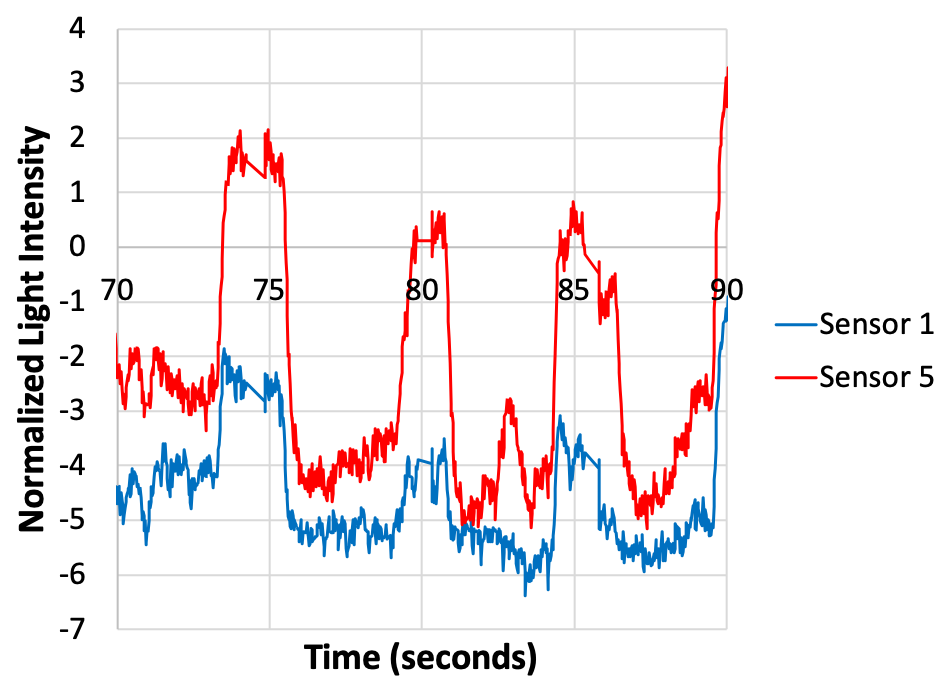}
  \caption{Normalized readings of sensors 1 and 5 for the ONE gesture }
  \label{fig:one15}
\end{figure}

\begin{figure}[t]
  \centering
  \includegraphics[width=\linewidth]{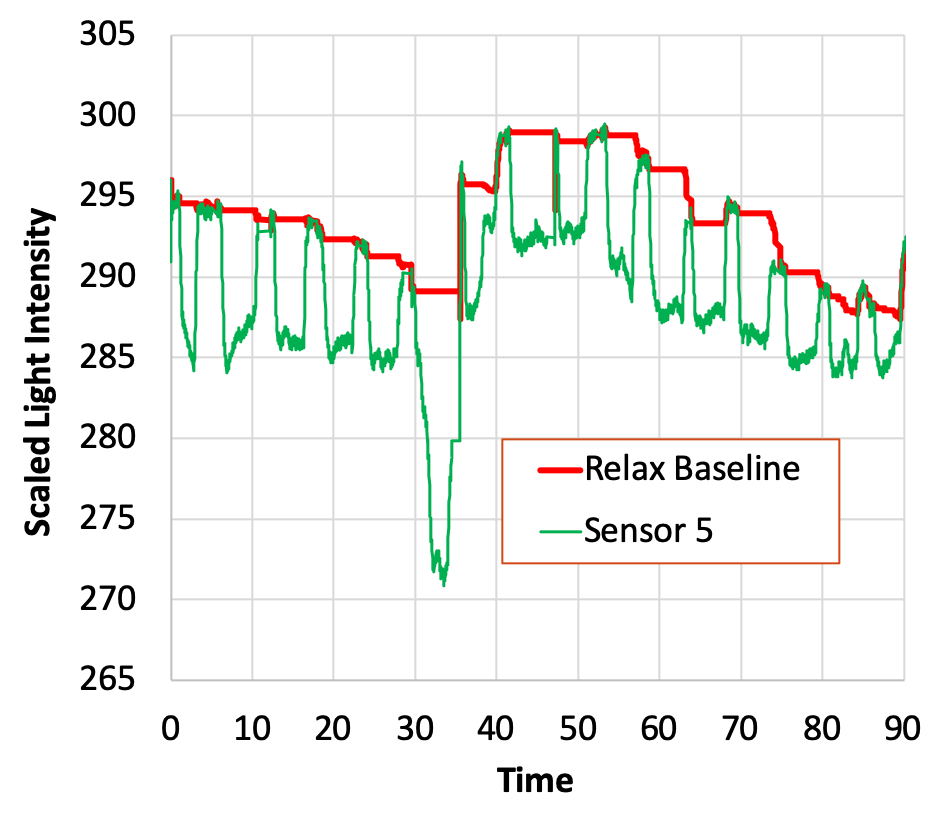}
  \caption{Readings of sensor 5 for the ONE gesture }
  \label{fig:one5}
\end{figure}

Figure~\ref{fig:one} plots the scaled light intensity of each sensor over time as I repeatedly performed a one finger gesture (ONE). There are several key takeaways from this experiment. First, each sensor (LDR) has its own unique baseline intensity value for the relaxed hand position. Second, each sensor reports a predictable change in intensity when the hand gestures are made. 

Figure~\ref{fig:one15} examines just the readings of sensors 1 and 5 for a portion of the experiment. This illustrates a third observation: the change due to the hand gesture differs across the different sensors. In this case, sensor 5's readings are impacted more significantly by this type of gesture. 

Figure~\ref{fig:one5} examines just the readings of sensor 5 for the above experiment. On this scale, the graph leads to the fourth observation:   while the change due to the hand gesture stays relatively consistent over time, the baseline reading for the relaxed hand position (the top plateaus of the green line) change significantly over time. I conjecture that this is due to movement or adjustment of the hand strap over periods longer than a single gesture. 

\subsection{Signature-Based Gesture Recognition}
\label{sec:signature}

Based on the above observations, I developed a gesture recognition algorithm with two key components. The first part establishes a moving baseline reading for the relax position. The second part compares the differences between the current reading and the baseline with \textit{signatures} created for each gesture to determine the current gesture. 

\subsubsection{Algorithm}

\paragraph{Baseline Computation.} The baseline value for the relax position is computed using two key techniques. First, I use an exponentially-weighted moving average of the sensor reading to remove any outliers and smooth noisy data, where: \[average_t = (1 - \alpha) \cdot average_{t-1} + \alpha \cdot X_t\] 
The $\alpha$ values determines how much "history" to include in the average -- small $\alpha$ values (close to 0) weigh the past more heavily causing the average to move more slowly with changes, while large $\alpha$ values (close to 1) cause the average to track the current readings more closely. Empirically, tuning $\alpha$ to 0.2 worked well at smoothing the sensor readings. In addition, I keep a \textit{sliding window} of the maximum smoothed average reading in the past 4 seconds. Gestures cause sensor values to dip. As a result, the maximum value over the recent past represents a good estimate of the current relax baseline reading. The red line in Figure~\ref{fig:one5} shows the result of the baseline computation. As the graphs shows, the baseline reading follows the changing value of the relax position extremely closely. 

\paragraph{Signature Generation}

\begin{figure}[t]
  \centering
  \includegraphics[width=\linewidth]{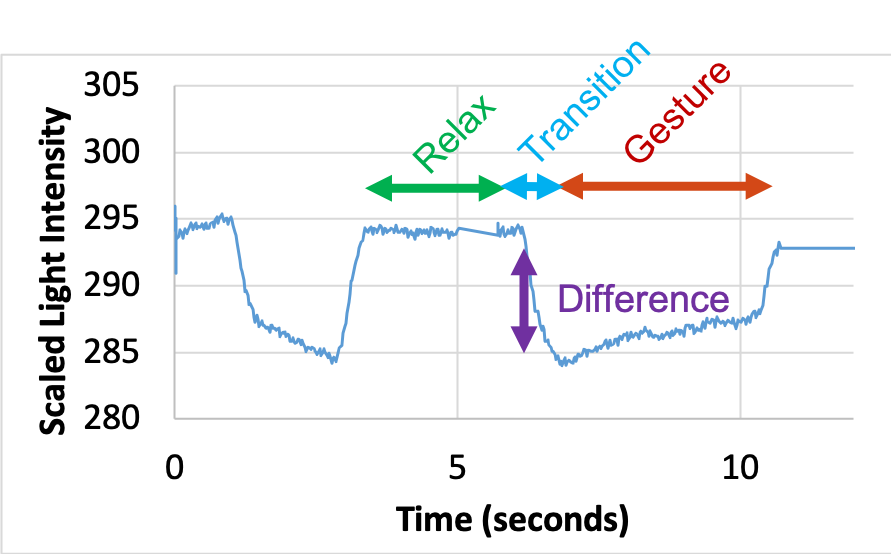}
  \caption{Labelling for signature generation.}
  \label{fig:sig1}
\end{figure}

\begin{figure}[t]
  \centering
  \includegraphics[width=\linewidth]{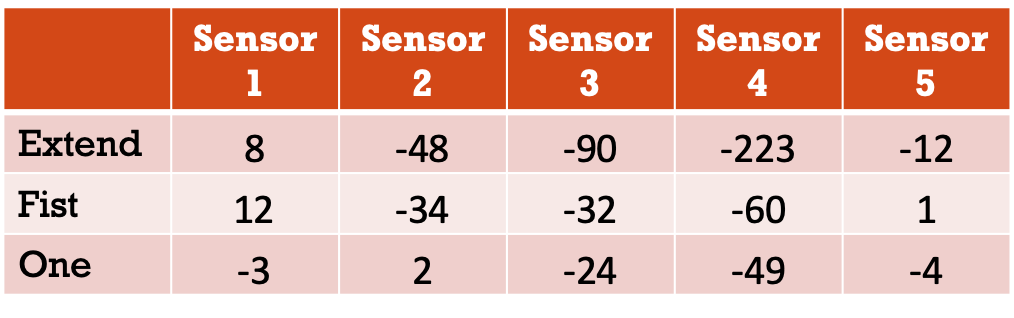}
  \caption{Signature values for gestures }
  \label{fig:sig2}
\end{figure}

For each dataset, I manually labelled the data readings of a single trace as “Relax Baseline”, “Transition Between Relax and Gesture” and “Gesture” (shown in Figure~\ref{fig:sig1}). For each section (e.g. a single Relax period) of the dataset, I computed a simple average of the readings for a single sensor. I then computed the difference in Light Intensity between a Relax period and the subsequent Gesture period. This was done for each sensor and for each gesture. For each gesture and sensor pair, the average of these difference values was computed to create a signature. Figure~\ref{fig:sig2} shows the signatures for the three gestures I am testing. The value $-223$ in the table indicates that the Sensor 4 reading typically dips by about $223$ from the baseline reading when the Extend gesture is performed. The table also shows that Sensors 2, 3 and 4 (i.e. the shorter fibers parallel to the arm) are impacted the most by the gestures. As a result, I use only these sensors for the gesture recognition in this algorithm. 

\paragraph{Gesture Identification}
To identify the current gesture, LiTe determines if the difference of the current reading from the baseline is within 10 percent of the signature value and, if so, it is labeled with that gesture. If the current reading is not within 10 percent of any label, it is identified as a RELAX gesture (i.e. baseline gesture). If the computed label had changed within the past 10 readings, the label was changed to TRANSITION.

\subsubsection{Accuracy}

To evaluate the accuracy of this algorithm, I used a dataset with a mix of all three gestures. I used the signatures described above to automatically compute labels for each data point. 

\begin{figure}[t]
  \centering
  \includegraphics[width=\linewidth]{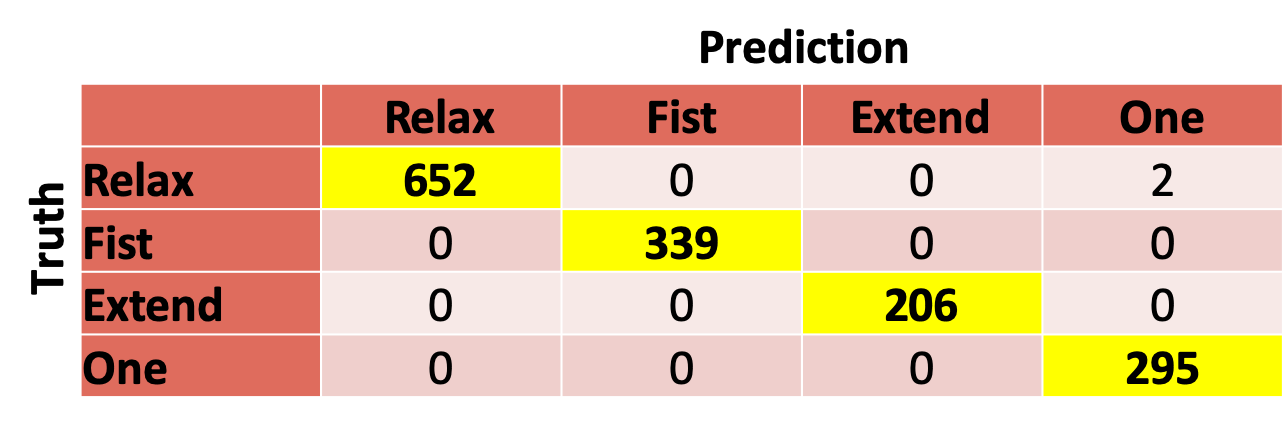}
  \caption{Confusion Matrix for Gesture Sensor Reading Signatures}
  \label{fig:confusion1}
\end{figure}

Accuracy of a classification system is typically summarized using a confusion matrix. A confusion matrix is a table used to describe the performance of a classification model (or "classifier") on a set of test data for which the true values are known.
 Each entry in the table indicates how often the true value in a row was predicted as the value in the column. The confusion matrix for the Gesture-based Gesture Classifier is shown in Figure~\ref{fig:confusion1}. Looking at the first row indicates that of the 654 instances of the Relax gesture, 652 were correctly identified as Relax and 2 were misclassified as One. We see that the bulk of the computed classifications are correct (highlighted along the diagonal). I omit the transition state since these would be filtered away in practice. Overall, only 2 of the 1494 readings were misclassified - i.e. an accuracy of 99.8\%.

\subsection{Machine Learning}
\label{sec:ML}

As an alternate approach, I evaluated using a neural network to classify the sensor readings to the different potential gestures. The benefit of a neural network is that it can automatically learn the relationship between the sensor values and the gestures - making it easier to add sensors or gestures to the system. 

\subsubsection{Algorithm}

Figure~\ref{fig:neuralnet} shows the Google Colaboratory~\cite{colaboratory} Python code to train this neural network. Lines 1--6 simply import the appropriate libraries and Lines 8--12 load a dataset containing a mix of sensor readings and their associated gesture labels. Lines 14--18 split the dataset into roughly two equal halves. The first half is for training the neural network and the second is for testing the trained model. 

Lines 20--32 construct the neural network. The neural network uses a Multi Layer Preceptron (MLP) structure with two hidden layers. This is well suited for learning a direct mapping of sensor values to output gesture types. However, unlike the manually created classification algorithm (Section~\ref{sec:software}), this approach does not consider the history of recent readings. I plan to evaluate the use of Recurrent Neural Networks (RNNs) and Long Short Term Memory Networks (LSTMs) to incorporate the use of recent readings as part of the gesture classification. 

Finally, lines 34-35 train the model using the training portion of the dataset. Lines 37-40 generate predictions using the trained model on the test portion of the dataset. To observe the accuracy of these predictions, the program outputs the confusion matrix between these predictions and the ground truth data. 

\begin{figure}[t]
\begin{lstlisting}[language=Python]
import numpy as np
import pandas as pd
import keras
from keras.models import Sequential
from keras.layers import Dense,Dropout
from keras.regularizers import l2

datafile = "mix6ml"
x_data = pd.read_csv(datafile+"X.csv",names=['s1','s2','s3','s4','s5'])
y_dataR = np.genfromtxt(datafile+"Y.csv",delimiter=',',dtype=int)
num_classes = np.max(y_dataR) + 1
y_data = keras.utils.to_categorical(y_dataR, num_classes)

# split the data into testing and training parts
x_train = x_data[:880]
x_test = x_data[880:]
y_train = y_data[:880]
y_test = y_data[880:]

#define a sequential Model
model = Sequential()

#Hidden Layer-1
model.add(Dense(100,activation='relu',input_dim=5,kernel_regularizer=l2(0.01)))
model.add(Dropout(0.3, noise_shape=None, seed=None))

#Hidden Layer-2
model.add(Dense(100,activation = 'relu',kernel_regularizer=l2(0.01)))
model.add(Dropout(0.3, noise_shape=None, seed=None))

#Output layer
model.add(Dense(6,activation='sigmoid'))

model.compile(loss='binary_crossentropy',optimizer='adam',metrics=['accuracy'])
model_output = model.fit(x_train,y_train,epochs=150,batch_size=20,verbose=1,validation_data=(x_test,y_test),)

y_pred = model.predict(x_test)
rounded = [np.argmax(x) for x in y_pred]
y_truth = [np.argmax(x) for x in y_test]
confusion_matrix(y_truth,rounded)
 \end{lstlisting}
  \caption{Neural network gesture classifier.}
  \label{fig:neuralnet}
\end{figure}
 
 \begin{figure}[t]
  \centering
  \includegraphics[width=\linewidth]{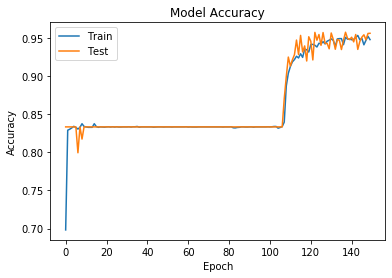}
  \caption{Accuracy during training}
  \label{fig:accuracy}
\end{figure}
 
  \begin{figure}[t]
  \centering
  \includegraphics[width=\linewidth]{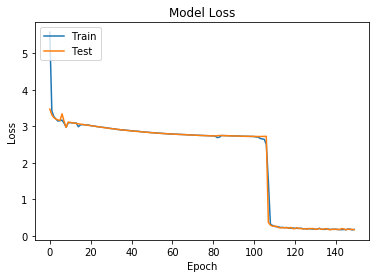}
  \caption{Loss during training}
  \label{fig:loss}
\end{figure}
 
 Figure~\ref{fig:accuracy} and \ref{fig:loss} show the performance of the trained model across the epochs of training. As can be seen in the accuracy graph, the final accuracy is relatively high rate (\char`\~ 0.95) at which the correct gestures are predicted on the training and test data. Similarly, the loss is also close to 0 by the end of training.  

  \begin{figure}[t]
  \centering
  \includegraphics[width=\linewidth]{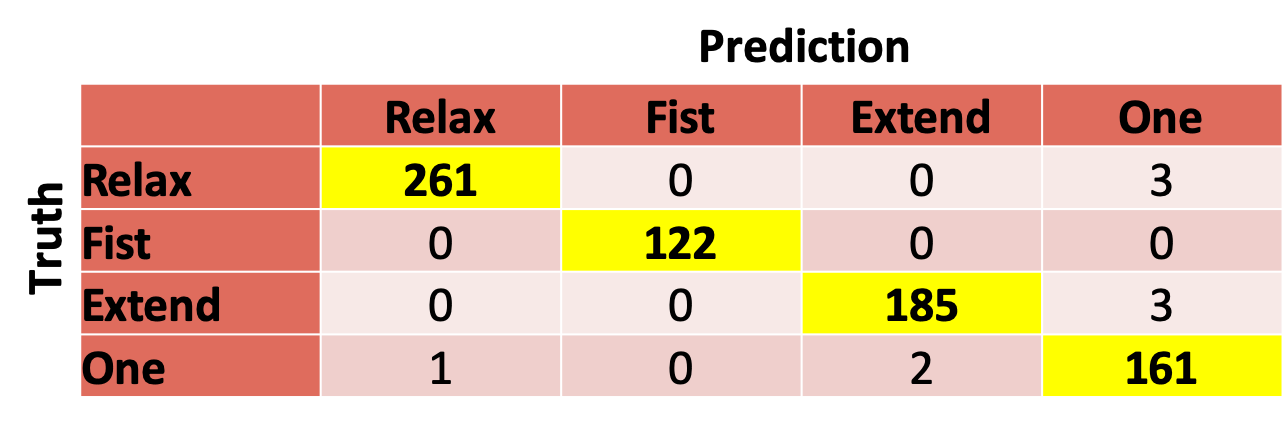}
  \caption{Confusion Matrix Created by Neural Network}
  \label{fig:confusion2}
\end{figure}

\subsubsection{Accuracy}
I use the trained model to predict the gestures for the test data readings. 
Figure~\ref{fig:confusion2} shows the confusion matrix for the three gestures and the relaxed hand state for the resulting prediction. I omit the transition state since these would be filtered away in practice.  Each entry in the table indicates how often the true value in a row was predicted as the value in the column. For example, of the 264 readings that were of the relaxed hand state, the neural network correctly predicted in 261 readings were relax and misclassified 3 of the readings as a \textit{one} gesture. The diagonal all represent correct predictions and show extremely high accuracy (98.8\%) for the neural network. The neural network does not perform quite as well as the manually created algorithm. I believe this is due to the fact that the Signature-based classifier incorporates history of past readings via the moving average computation to account for changing conditions. It may be possible to use an alternate neural network architecture that explicitly incorporates history to improve the system's performance further. However, with such a high accuracy using the current approach this optimization may prove unnecessary. 

\section{Discussion}
\label{sec:discussion}

The experimental study in the previous section enables me to make the following important observations: Optical fibers do make feasible sensors for detecting gestures. It is possible to distinguish between gestures with almost 99 percent accuracy. It is also possible to use neural networks to automate the gesture recognition process.

While these results show promise in the approach of using custom optical sensors to detect the deflection of tendons in the wrist, there are limitations in research that must be considered in making this approach practical.


\paragraph{Variations in tendons} Testing was conducted only on the author himself. As a result, it is possible that there are human variations in how the gestures are made or how the tendons move for a different user.

\paragraph{Realism} The prototype is powered by an Arduino. Integrating LiTe into a smart watch would take another step.

\paragraph{Sign Language} Sign language involves more than just the movement of individual fingers. American Sign Language (ASL) requires movement of the entire hand and arm. The accuracy of the system to detect ASL and not just simple gestures will need to be continued to explore.

\section{Conclusion and Next Steps}

My study shows that fiber optics are, indeed, a practical and light-weight method of detecting deflection in the wrist. This method is inexpensive compared to all existing technologies. The prototype created in this project was less than \$30, hence also achieving the low-cost criteria. The results of the project demonstrate that it is promising first step in creating a light-weight assistive technology for the speech and hearing impaired. 

The next step of the project is expand the number of gestures and continue to process them using machine learning. An eventual goal is to incorporate the entire system into an existing watch band and have it powered by a smart watch rather than an Arduino.

\section*{Acknowledgements}

Professor Mayank Goel and the smashlab (\cite{smashlab}) at Carnegie Mellon University provided a place to conduct the research. The design, prototypes, 3D printed models, and molds were all created by the author of this paper. The molding process was completed in the lab under the guidance of a smashlab graduate student in the summer of 2019. The materials for fabrication were purchased and provided by the lab. The entirety of the data collection, data analysis, and machine learning components were conducted independently by the author in the fall and winter of 2019.


\bibliographystyle{ACM-Reference-Format}
\bibliography{_references}


\begin{thebibliography}{12}


\ifx \showCODEN    \undefined \def \showCODEN     #1{\unskip}     \fi
\ifx \showDOI      \undefined \def \showDOI       #1{#1}\fi
\ifx \showISBNx    \undefined \def \showISBNx     #1{\unskip}     \fi
\ifx \showISBNxiii \undefined \def \showISBNxiii  #1{\unskip}     \fi
\ifx \showISSN     \undefined \def \showISSN      #1{\unskip}     \fi
\ifx \showLCCN     \undefined \def \showLCCN      #1{\unskip}     \fi
\ifx \shownote     \undefined \def \shownote      #1{#1}          \fi
\ifx \showarticletitle \undefined \def \showarticletitle #1{#1}   \fi
\ifx \showURL      \undefined \def \showURL       {\relax}        \fi
\providecommand\bibfield[2]{#2}
\providecommand\bibinfo[2]{#2}
\providecommand\natexlab[1]{#1}
\providecommand\showeprint[2][]{arXiv:#2}

\bibitem[\protect\citeauthoryear{??}{cyb}{2020}]%
        {cyberglove}
 \bibinfo{year}{2020}\natexlab{}.
\newblock \bibinfo{title}{Cyberglove}.
\newblock
\newblock
\urldef\tempurl%
\url{https://developer.microsoft.com/en-us/windows/kinect/}
\showURL{%
\tempurl}


\bibitem[\protect\citeauthoryear{??}{col}{2020}]%
        {colaboratory}
 \bibinfo{year}{2020}\natexlab{}.
\newblock \bibinfo{title}{Google Colaboratory}.
\newblock
\newblock
\urldef\tempurl%
\url{https://colab.research.google.com/notebooks/intro.ipynb/}
\showURL{%
\tempurl}


\bibitem[\protect\citeauthoryear{??}{lea}{2020}]%
        {leap}
 \bibinfo{year}{2020}\natexlab{}.
\newblock \bibinfo{title}{LEAP Motion Control}.
\newblock
\newblock
\urldef\tempurl%
\url{https://www.ultraleap.com/product/leap-motion-controller/}
\showURL{%
\tempurl}


\bibitem[\protect\citeauthoryear{??}{kin}{2020}]%
        {kinect}
 \bibinfo{year}{2020}\natexlab{}.
\newblock \bibinfo{title}{Microsoft Kinect}.
\newblock
\newblock
\urldef\tempurl%
\url{http://www.cyberglovesystems.com}
\showURL{%
\tempurl}


\bibitem[\protect\citeauthoryear{??}{sma}{2020}]%
        {smashlab}
 \bibinfo{year}{2020}\natexlab{}.
\newblock \bibinfo{title}{Smart Sensing For Humans}.
\newblock
\newblock
\urldef\tempurl%
\url{http://smashlab.io/}
\showURL{%
\tempurl}


\bibitem[\protect\citeauthoryear{??}{tap}{2020}]%
        {tap}
 \bibinfo{year}{2020}\natexlab{}.
\newblock \bibinfo{title}{TAP}.
\newblock
\newblock
\urldef\tempurl%
\url{https://www.tapwithus.com}
\showURL{%
\tempurl}


\bibitem[\protect\citeauthoryear{??}{pix}{Feb}]%
        {pixel4}
 \bibinfo{year}{Feb.}\natexlab{}.
\newblock \bibinfo{title}{Google Pixel 4}.
\newblock
\newblock
\urldef\tempurl%
\url{https://store.google.com/product/pixel_4_specs}
\showURL{%
\tempurl}


\bibitem[\protect\citeauthoryear{Boundless}{Boundless}{[n. d.]}]%
        {physics}
\bibfield{author}{\bibinfo{person}{Boundless}.} \bibinfo{year}{[n.
  d.]}\natexlab{}.
\newblock \bibinfo{title}{Total Internal Reflection and Fiber Optics}.
\newblock
\newblock
\urldef\tempurl%
\url{http://oer2go.org/mods/en-boundless/www.boundless.com/physics/textbooks/boundless-physics-textbook/geometric-optics-24/reflection-refraction-and-dispersion-169/total-internal-reflection-and-fiber-optics-609-6258/index.html}
\showURL{%
\tempurl}


\bibitem[\protect\citeauthoryear{C.~To and Park}{C.~To and Park}{2015}]%
        {strechable}
\bibfield{author}{\bibinfo{person}{T.~L.~Hellebrekers C.~To} {and}
  \bibinfo{person}{Y. Park}.} \bibinfo{year}{2015}\natexlab{}.
\newblock \showarticletitle{Highly stretchable optical sensors for pressure,
  strain, and curvature measurement}.
\newblock  (\bibinfo{year}{2015}).
\newblock


\bibitem[\protect\citeauthoryear{Foundation}{Foundation}{2020}]%
        {biology}
\bibfield{author}{\bibinfo{person}{Arthritis Foundation}.}
  \bibinfo{year}{2020}\natexlab{}.
\newblock \bibinfo{title}{Hand and Wrist Anatomy}.
\newblock
\newblock
\urldef\tempurl%
\url{https://www.arthritis.org/health-wellness/about-arthritis/where-it-hurts/hand-and-wrist-anatomy}
\showURL{%
\tempurl}


\bibitem[\protect\citeauthoryear{Van~Meerbeek, De~Sa, and
  Shepherd}{Van~Meerbeek et~al\mbox{.}}{2018}]%
        {optoelectronic}
\bibfield{author}{\bibinfo{person}{I.~M. Van~Meerbeek}, \bibinfo{person}{C.~M.
  De~Sa}, {and} \bibinfo{person}{R.~F. Shepherd}.}
  \bibinfo{year}{2018}\natexlab{}.
\newblock \showarticletitle{Soft optoelectronic sensory foams with
  proprioception}.
\newblock \bibinfo{journal}{\emph{Science Robotics}} \bibinfo{volume}{3},
  \bibinfo{number}{24} (\bibinfo{year}{2018}).
\newblock
\urldef\tempurl%
\url{https://doi.org/10.1126/scirobotics.aau2489}
\showDOI{\tempurl}
\showeprint{https://robotics.sciencemag.org/content/3/24/eaau2489.full.pdf}


\bibitem[\protect\citeauthoryear{Weissbrod}{Weissbrod}{2020}]%
        {medicalart}
\bibfield{author}{\bibinfo{person}{Elizabeth Weissbrod}.}
  \bibinfo{year}{2020}\natexlab{}.
\newblock \bibinfo{title}{Muscles and Tendons in the Hand}.
\newblock
\newblock
\urldef\tempurl%
\url{https://medicalart.johnshopkins.edu/portfolio-item/muscles-and-tendons-in-the-hand/}
\showURL{%
\tempurl}


\end{thebibliography}
\end{document}